\documentclass[preprintnumbers,amsmath,amssymb,floatfix,11pt,prd,onecolumn,superscriptaddress,nofootinbib]{revtex4}
\usepackage{amsmath,amsfonts,dsfont,amsthm,amssymb}
\usepackage[colorlinks=true, pdfstartview=FitV, linkcolor=blue, citecolor=red, urlcolor=magenta]{hyperref}
\usepackage{caption}
\usepackage{mathtools}
\usepackage{pdflscape}
\usepackage{multirow}
\usepackage{graphicx}
\usepackage{color}
\usepackage{epsfig}
\usepackage{verbatim}
\usepackage{float}
\usepackage{hyperref}
\theoremstyle{plain}

\theoremstyle{definition}

\usepackage{subfigure}

\theoremstyle{remark}

\numberwithin{equation}{section}

\begin{document}
\title{\bf Repetitive Penrose Process in Rastall Rotating Black Holes Immersed in Quintessence Dark Energy}

\author{Ali Ahmad Sabir}
\altaffiliation{aasabir23@gmail.com}
\affiliation{Department of Mathematics, University of Okara, Okara-56300 Pakistan}

\author{Muhammad Israr Aslam}
\altaffiliation{mrisraraslam@gmail.com, israr.aslam@umt.edu.pk}
\affiliation{Department of Mathematics, School of Science, University of Management and Technology, Lahore-$54770$, Pakistan.}

\author{Abdul Malik Sultan}
\altaffiliation{ams@uo.edu.pk, maliksultan23@gmail.com}
\affiliation{Department of Mathematics, University of Okara, Okara-56300 Pakistan}

\author{Ke Wang}
\altaffiliation{kkwwang2025@163.com}
\affiliation{School of Material Science and Engineering, Chongqing Jiaotong University, Chongqing 400074, China}

\author{Hamood Ur Rehman}
\altaffiliation{hamood@uo.edu.pk}
\affiliation{Department of Mathematics, University of Okara, Okara-56300 Pakistan}

\author{Yakup Yildirim}
\altaffiliation{yyildirim@biruni.edu.tr}
\affiliation{Department of Computer Engineering, Biruni University, Istanbul–34010, Turkey.}

\begin{abstract}
We investigate the repetitive Penrose process in the spacetime of a Rastall rotating black hole surrounded by a quintessence dark energy field. After reviewing the fundamental properties of the black hole geometry, we formulate the repetitive Penrose process by deriving the conservation equations governing particle splitting within the ergoregion, along with the corresponding iterative evolution equations. The physical conditions required for terminating the energy extraction iterations are established, and the minimum spin thresholds of the decay particles are analyzed to identify the critical stopping criterion. Our analysis reveals that the termination of the repetitive Penrose process is consistently governed by Particle~$0$, which possesses the highest minimum spin threshold among all decay products. Numerical results further demonstrate that the dimensionless Rastall structure parameter $\hat{N}_s$ and the Rastall coupling parameter $\alpha$ significantly influence the evolution of the energy extraction process. At the same decay radii increasing initial values of both parameters boosts the energy utilization efficiency and energy return on investment. Specifically, smaller values of  $\hat{N}_s$ enhances the energy utilization efficiency at lower decay radii, shifts the maximum extracted energy toward lower decay radii, and accelerates the depletion of the remaining extractable energy reservoir. This indicates that the repetitive Penrose process is highly favored at lower decay radii. Smaller initial values of $\hat{N}_s$ yield a larger maximum energy return on investment. Similarly, increasing $\alpha$ enhances the energy utilization efficiency, alters the location of the peak extracted energy, and reduces the total extractable energy. But the effects of $\alpha$ on these energetics are very small as compared to $\hat{N}_s$. These findings highlight the pivotal role of Rastall gravity and quintessence dark energy in shaping the efficiency and  dynamics of black hole rotational energy extraction.
\end{abstract}
\date{\today}
\maketitle

\section{Introduction}\label{intro}
Black holes are among the most remarkable predictions of Einstein's theory of general relativity (GR) and play a central role in modern relativistic astrophysics. A wide range of high-energy astrophysical phenomena, such as active galactic nuclei, relativistic jets, quasars, gamma-ray bursts, and X-ray binaries, reveal that rotating black holes act as extremely efficient astrophysical energy sources in the Universe  \cite{McKinney2004,Hawley2006,Tchekhovskoy2011,Lee2000,Penrose1969}. Understanding the physical mechanisms responsible for extracting rotational energy from black holes is therefore of considerable theoretical and observational interest. In particular, the interaction between strong gravitational fields and surrounding matter provides a natural environment for investigating how rotational energy can be converted into observable high-energy phenomena. Such studies not only improve our understanding of compact-object dynamics but also offer valuable insights into the behavior of matter and radiation in the strong-field regime of gravity. Consequently, the investigation of energy extraction processes has become an important link between theoretical models of black holes and their observable astrophysical signatures.

Rotating black holes possess enormous reservoirs of rotational energy that can be extracted through different physical mechanisms. Among these, a process was established by Penrose, who showed that a particle entering the ergosphere splits into two fragments, where one fragment falls into the black hole with negative energy while the other escapes to infinity carrying more energy than the original particle \cite{RPenrose1971}. This mechanism directly utilizes the existence of negative energy states inside the ergoregion of a rotating spacetime. The Penrose process has therefore become a fundamental theoretical framework for investigating the conversion of black hole rotational energy into escaping particle energy and has inspired numerous studies of relativistic energy extraction. Its underlying principles have also motivated the development of several related mechanisms operating in magnetized and plasma-filled environments surrounding astrophysical black holes. However, subsequent investigations showed that the original Penrose process requires extremely high relative velocities between the decay fragments, making its direct astrophysical realization difficult \cite{Wald1974,Piran1975}. Consequently, considerable effort has been devoted to exploring alternative scenarios and refinements that overcome these limitations while preserving the basic concept of rotational energy extraction.

The Penrose mechanism has motivated extensive research into black hole energy extraction processes, providing the theoretical framework for various mechanisms through which the rotational energy of spinning black holes can be investigated. Several important extensions have been developed, including collisional Penrose processes \cite{Banados2009,Blandford1977}, electromagnetic extraction mechanisms such as the Blandford-Znajek process \cite{Takahashi1990}, magnetic reconnection models \cite{Comisso2021} and generalized Penrose processes in modified gravitational backgrounds \cite{Dhurandhar1984a,Dhurandhar1984b}. Consequently, the transformation of the rotational energy of a Kerr black hole has been investigated in \cite{china1}. Recently, Ruffini et al. \cite{Ruffini2025a} have revisited the Penrose process by introducing turning-point conditions for the particle trajectories and have proposed the repetitive Penrose process, in which the extraction procedure is repeated after each decay event. Their analysis demonstrated that not all rotational energy of the black hole can be extracted because a significant fraction of the energy variation contributes to the increase of the irreducible mass. The repetitive Penrose process has subsequently been extended to charged and non-asymptotically flat black hole spacetimes, including the repetitive electroprocess, Kerr-de Sitter black holes, accelerating Kerr black holes, Kerr-Taub-NUT black holes, and charged particles in Kerr-Newman black holes \cite{Lhu2026,Wang2025,Zeng2026,Zeng2,china2,china3}. More recently, Alipour et al. \cite{guass} investigated the repetitive Penrose process in rotating four-dimensional Einstein-Gauss-Bonnet black holes.

The theory of GR, as the standard theory of classical gravity, has obtained remarkable achievements on solar system scales and in gravitational wave observations. However, from a theoretical point of view, GR suffers from non-renormalizability in the ultraviolet regime, indicating that it needs to be replaced by an ultraviolet-complete quantum theory of gravity at extreme energy scales. From an observational perspective, the discovery of the late-time accelerated expansion of the Universe, together with the inflationary epoch in the early Universe, has provided strong motivation for exploring extensions and modifications of GR. In this context, among the pool and the proposal to extend GR, a compelling modification was proposed by Rastall
in 1972 \cite{Rastall1972}. Specifically, this theory provides interesting modifications in which the usual conservation of energy is generalized according to $\nabla_{\mu}T^{\mu\nu}=\alpha \nabla^{\nu}R$, where $\alpha$ denotes the Rastall coupling parameter and $R$ is the Ricci scalar \cite{Rastall1972}. In this framework, the non-minimal coupling between geometry and matter modifies the spacetime dynamics in curved backgrounds while recovering the standard conservation law in the appropriate limit. The investigation of cosmological model within the fabric of Rastall gravity along with its consistency with various cosmic eras are analyzed in \cite{rastall1}. Additionally, Visser and Darabi have presented an important arguments regarding the consistency and physical interpretation of Rastall gravity, specifically concerning its compatibility with GR (for a detail review one can see Refs. \cite{Visser2018,rastall2}). Subsequently, Rastall gravity has admitted all the challenges from cosmological as well as astrophysical measurements \cite{Lin2019,rastall3,rastall4}. Recently, several researchers have carried out extensive investigations into various aspects of Rastall gravity, including black hole solutions \cite{Heydarzade2017,rastall5,rastall6,YHeydarzade2017,rastall7}, wormhole geometries \cite{rastall8,rastall9}, cosmological evolution \cite{rastall10,rastall11} and many more found in literature. Moreover, the thermodynamical quantities and visual signatures, including the shadows of Rastall rotating black holes, have been extensively investigated in recent studies \cite{Rkumar2018,Rkumar2021,AMSultan2026}.

Rotating black hole solutions in Rastall gravity provide a useful framework for examining the influence of modified gravitational dynamics on horizon geometry and particle \cite{2Lin2019}. Specifically, the horizon and ergosphere structures of a rotating Rastall black hole are significantly influenced by the Rastall parameter and the properties of the surrounding matter  \cite{rastallnew1}. Since the efficiency of the Penrose process is closely related to the geometry of the ergoregion, significant modifications in the energy extraction mechanism may be expected in comparison with the Kerr spacetime \cite{Penrose1969,RPenrose1971,Wald1974}. These geometric changes can alter the conditions for negative-energy particle trajectories inside the ergoregion and consequently affect the amount of rotational energy that can be extracted. Investigating these effects is therefore essential for assessing how deviations from GR influence classical energy extraction mechanisms. Nevertheless, the influence of these modified spacetime features on repetitive energy extraction processes remains largely unexplored, leaving several important aspects of the Penrose mechanism unexplored \cite{Lin2019}.

Despite these investigations, the repetitive Penrose process in the Rastall rotating black hole spacetime has not yet been studied in detail. In particular, the influence of the Rastall parameter on the iterative extraction of rotational energy, the evolution of the irreducible mass, and the stopping conditions of the repetitive Penrose process remain unclear. Studying these effects is important for understanding the interplay between modified gravity and black hole energetics. A systematic investigation can reveal how the non-minimal coupling between matter and geometry in Rastall gravity influences the efficiency and long-term evolution of the extraction process compared with the predictions of GR.  Such an analysis also provides a deeper understanding of the role of spacetime geometry in determining the ultimate limits of rotational energy extraction. These results may further clarify whether Rastall gravity leaves distinguishable signatures in black hole energetics that could be explored in future theoretical and astrophysical studies.

Motivated by the aforementioned literature, in this work we investigate the repetitive Penrose process in the spacetime of a Rastall rotating black hole surrounded by a quintessence field. We specifically investigate the role of the Rastall structure parameter and the Rastall coupling parameter in determining the amount of iteratively extractable energy, the evolution of the black hole irreducible mass, and the termination conditions of the energy extraction process. We investigate both analytical and numerical analyses of how these parameters affect the investment return of energy, energy extraction efficiency, the extracted energy, and the remaining extractable energy. By extending the recently proposed repetitive Penrose process from the Kerr spacetime to a more general Rastall gravitational background, we perform a general analysis of the rotational energy extraction within the fabric of the Rastall theory of gravity.

This manuscript is organized as follows. In section \ref{sec2}, we will define the background of the Penrose process within the framework of Rastall rotating black hole. In section \ref{sec3}, we will the discuss the iterative stopping conditions for the repetitive energy extraction process. In section \ref{sec4}, we will investigate the dynamics of the repetitive Penrose process, focusing specifically on how the Rastall structure parameter $\hat{N}_s$ and Rastall coupling parameter $\alpha$ influence on this mechanism. We will summarize our main results in section \ref{sec5}. Throughout this study, we will employ geometric units $c = G = 1$.

\section{The Penrose Process in Rastall Rotating Black Holes}\label{sec2}
The metric for rotating
Rastall black hole in the Boyer-Lindquist coordinates is defined as \cite{Rkumar2018}
\begin{eqnarray}\nonumber{}
ds^2&=&-\left(1-\frac{2Mr + N_s r^\zeta}{\Sigma}\right)dt^2
-\frac{2a\sin^2\theta\left(2Mr+N_s r^\zeta\right)}{\Sigma}dt\,d\phi +\frac{\Sigma}{\Delta}dr^2 +\Sigma d\theta^2 \\\label{1} &+& \sin^2\theta \left[ r^2 + a^2+ \frac{a^2\sin^2\theta\left(2Mr + N_s r^\zeta\right)}{\Sigma} \right]d\phi^2,
\end{eqnarray} 
where
\begin{eqnarray}
    \Sigma=r^2+a^2\cos^2{\theta} \quad \text{and} \quad \Delta=r^2-2Mr+a^2-N_sr^{\zeta},
\end{eqnarray}
in which $M$ is the black hole mass, $a$ corresponds to the spin parameter, $\zeta=\frac{(1-3\omega_s)}{1-3\alpha(1+\omega_s)}$ and $N_s$ is the surrounding field structure parameter. In the absence of a surrounding field, i.e., $N_{s}\rightarrow 0$, the corresponding metric reduces to the usual Kerr black hole \cite{rastallnew2}, while $\omega_s$ is the equation of state parameter of the surrounding fluid. Specifically, when $\alpha=0$ and $-1<\omega_s<\frac{-1}{3}$, the metric represents the Kerr black hole surrounded by a quintessence field \cite{rastallnew3}, while the Schwarzschild black hole solution can be recovered when both $N_{s}$ and $a$ approaches to zero. In the subsequent analysis, we fixed $\omega_s=\frac{-2}{3}$, which represents the metric surrounded by a quintessence field within the fabric of Rastall gravity.

The event horizon of the black hole is calculated by $\Delta=0$, and the boundary of the ergosphere is evaluated by $g_{tt}=0$. For a detailed discussion of the event horizon and the ergosphere, one can see Ref. \cite{Rkumar2018}. Unlike the standard Kerr spacetime, the addition of both Rastall field parameter $N_s$ and $\zeta$ significantly complicates the algebraic structure of the horizon function $\Delta$ and the temporal metric component  $g_{tt}$. Because these transcendental equations cannot be solved analytically, therefore one can evaluate the location of the event horizons and the boundary of the ergosphere numerically. Solving $\Delta = 0$ and $g_{tt} = 0$ yields multiple roots. In this analysis, we consider only the largest positive real roots, corresponding to the event horizon $r_+$ and the outer ergosphere radius $r_{E}$, which define the observable spacetime boundaries. For comparison with the Kerr black hole framework, we adjust the initial spin of the black hole to $a=M$ and consider the particle motion in the equatorial plane. The region where the particle splitting process takes place, which is the ergosphere, lies between $r_+$ and $r_E$. The surface area of the event horizon of the Rastall rotating black hole is defined as
\begin{equation}
S=\int_{0}^{\pi}\int_{0}^{2\pi}
\sqrt{g_{\theta\theta}g_{\phi\phi}}
\, d\phi\, d\theta
=4\pi\left(r_{+}^{2}+a^{2}\right),
\end{equation}
where $r_+$ denotes the outer event horizon determined from $\Delta(r_+)=0$. Using the Bekenstein--Hawking area law, the irreducible mass of the black hole is given by \cite{Dchris}
\begin{equation}
M_{\rm irr}=\sqrt{\frac{S}{16\pi}}
=\frac{1}{2}\sqrt{r_{+}^{2}+a^{2}}.
\end{equation}
Consequently, the maximum rotational energy that can be extracted from the black hole can be expressed as
\begin{equation}
E_{\rm extractable}=M-M_{\rm irr}=M-\frac{1}{2}\sqrt{r_{+}^{2}+a^{2}}.
\end{equation}
The above expressions are obtained under the assumption that the entropy of the Rastall rotating black hole obeys the standard
Bekenstein-Hawking area law, according to which the entropy is proportional to the area of the event horizon. Unlike modified gravity theories containing explicit higher-order curvature corrections, where the entropy may acquire additional contributions, the Rastall framework modifies the spacetime geometry primarily through the horizon structure. Moreover, the Rastall coupling parameter $\alpha$ and the surrounding field parameter $N_s$ influence the thermodynamic quantities indirectly by shifting the position of the event horizon $r_+$. Consequently, the horizon area, irreducible mass, and amount of extractable energy differ from their Kerr counterparts through the corresponding change in the horizon radius. In the limits $N_{s} \rightarrow 0$ and $\alpha \rightarrow 0$, the Rastall rotating black hole smoothly approaches the Kerr black hole solution. Under these constraints, the expressions for the horizon area, irreducible mass, and extractable energy can be reduced to the well-known results of the Kerr spacetime.

\begin{figure}[htbp]
\centering
\begin{minipage}[b]{0.46\textwidth}
\includegraphics[width=\textwidth]{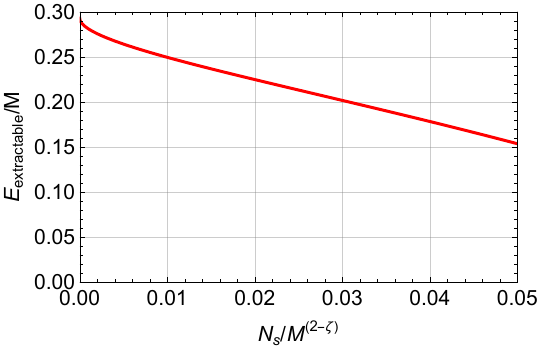}
\end{minipage}
\begin{minipage}[b]{0.46\textwidth}
\includegraphics[width=\textwidth]{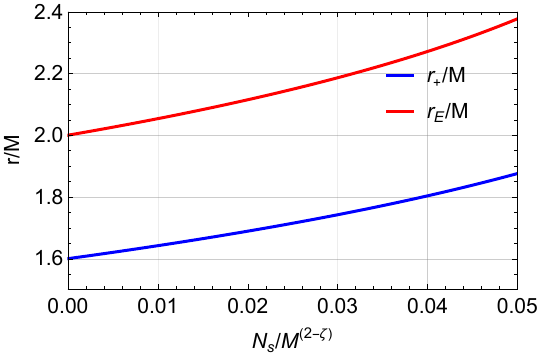}
\end{minipage}
\begin{minipage}[b]{0.46\textwidth}
\includegraphics[width=\textwidth]{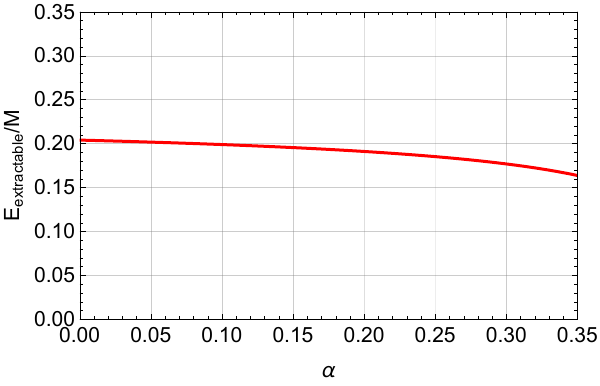}
\end{minipage}
\begin{minipage}[b]{0.46\textwidth}
\includegraphics[width=\textwidth]{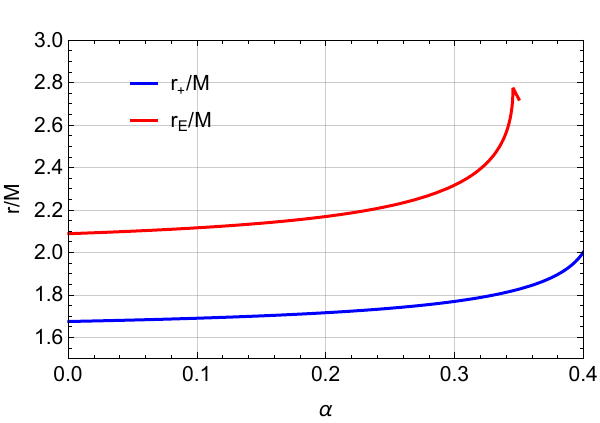}
\end{minipage}
\caption{Variation of $E_{extractable}/M$, $r_+/M$ and $r_E/M$, as functions of $\hat{N}_{s}$ (top row) with fixed $\alpha=0.1$, and as functions of $\alpha$ with fixed $\hat{N}_s=0.02$ (bottom row).}\label{fig1}
\end{figure}

In \textbf{FIG.}~\ref{fig1} we plot the maximum extractable energy, the event horizon, and the boundary of the ergosphere for $a=M$ as functions of the parameters $\hat{N_{s}}=\frac{N_s}{M^{(2-\zeta)}}$ and $\alpha$. It can be seen that the extractable energy decreases with increasing both parameters and the size of the event horizon increases monotonically, maintaining a relatively stable and uniform gap between the two radii. In the Penrose mechanism, the decay of a particle within the ergoregion must satisfy the conservation of four-momentum. Consequently, the conservation relations for the energy, angular momentum, and radial momentum of the participating particles can be written as
\begin{eqnarray}
\hat{E}_{0}=\hat{\mu}_{1}\hat{E}_{1}+\hat{\mu}_{2}\hat{E}_{2}, \quad \hat{p}_{\phi 0}=\hat{\mu}_{1}\hat{p}_{\phi 1}+\hat{\mu}_{2}\hat{p}_{\phi 2}, \quad \hat{p}_{r0}=\hat{\mu}_{1}\hat{p}_{r1}+\hat{\mu}_{2}\hat{p}_{r2},
\end{eqnarray}\label{eq:energy_cons}
in which
\begin{equation}
\hat{\mu}_{i}=\frac{\mu_{i}}{\mu_{0}},
\qquad
\hat{E}_{i}=\frac{E_{i}}{\mu_{i}},
\qquad
\hat{p}_{\phi i}=\frac{p_{\phi i}}{\mu_{i}M},
\qquad
\hat{p}_{ri}=\frac{p_{ri}}{\mu_{i}},
\qquad
i\in 0,1,2.\label{eq:dimensionless}
\end{equation}
Here, $\mu_{i}$ denotes the rest mass of the $i$th particle. For convenience, all physical quantities are expressed in dimensionless form by scaling them with appropriate combinations of black hole mass $M$ and incident particle mass $\mu_0$. The maximum efficiency of the Penrose process is achieved when the particle decay occurs at a turning point of the radial motion. In this configuration, the radial momenta of the incident particle and the two resulting fragments vanish at the splitting location. Consequently, all three particles satisfy the turning-point condition $\hat{E}_i=\hat{V}_i^{+}$, and the effective potential for the equatorial motion is given by \cite{Wang2025}

\begin{equation}
\hat{V}_{i}^{\pm}=
\frac{g^{\phi t}\hat{p}_{\phi i}M
\mp
\sqrt{\left(g^{\phi t}\right)^2\hat{p}_{\phi i}^{\,2}M^2-g^{tt}
\left(
g^{\phi\phi}\hat{p}_{\phi i}^{\,2}M^2+1
\right)}}{g^{tt}}.\label{eq:effective_potential}
\end{equation}
Under these conditions, the conservation equations of the Penrose process admit analytical solutions. Assuming that the quantities $\hat{E}_0$, $\hat{p}_{\phi 1}$, and $\nu=\mu_2/\mu_1$ are specified, the fundamental equations of the Penrose process have an analytic solution \cite{Wang2025}
\begin{eqnarray}\label{eq:mus1}
\hat{p}_{\phi 0} &=&\frac{g^{\phi t}\hat{E}_0 + \sqrt{ \left(g^{\phi t}\right)^2\hat{E}_0^{\,2}-
g^{\phi\phi} \left( 1+g^{tt}\hat{E}_0^{\,2} \right)}}{M g^{\phi\phi}}, \\\label{eq:muq1} \hat{E}_1
&=&\frac{
g^{\phi t}\hat{p}_{\phi 1}M-\sqrt{\left(g^{\phi t}\right)^2\hat{p}_{\phi 1}^{\,2}M^2-g^{tt}\left(g^{\phi\phi}\hat{p}_{\phi 1}^{\,2}M^2+1\right)}}{g^{tt}},
\\\label{eq:mu1}
\hat{\mu}_1
&=&
\frac{\hat{E}_0\hat{E}_1g^{tt}-\hat{E}_1g^{\phi t}M\hat{p}_{\phi0}-\hat{E}_0g^{\phi t}M\hat{p}_{\phi1}+g^{\phi\phi}M^2\hat{p}_{\phi0}\hat{p}_{\phi1}+\sqrt{D}}
{\hat{E}_1^{\,2}g^{tt}
-2\hat{E}_1g^{\phi t}M\hat{p}_{\phi1}
+g^{\phi\phi}M^2\hat{p}_{\phi1}^{\,2}
+\nu^{2}},
\end{eqnarray}
and
\begin{equation}
\hat{p}_{\phi 2} = \frac{\hat{p}_{\phi 0}}{\tilde{\mu}_2} - \frac{\hat{p}_{\phi 1}}{\nu}, \quad 
\hat{E}_2 = \frac{\hat{E}_0}{\tilde{\mu}_2} - \frac{\hat{E}_1}{\nu},
\end{equation}
where
\begin{equation}
\begin{aligned}
D=& - g^{tt} g^{\phi\phi} M^2 \hat{E}_1^2 \hat{p}_{\phi 0}^2 + (g^{\phi t})^2 M^2 \hat{E}_1^2 \hat{p}_{\phi 0}^2 - g^{tt} g^{\phi\phi} M^2 \hat{E}_0^2 \hat{p}_{\phi 1}^2 + (g^{\phi t})^2 M^2 \hat{E}_0^2 \hat{p}_{\phi 1}^2 \\
& - 2 (g^{\phi t})^2 M^2 \hat{E}_0 \hat{E}_1 \hat{p}_{\phi 0} \hat{p}_{\phi 1} + 2 g^{tt} g^{\phi\phi} M^2 \hat{E}_0 \hat{E}_1 \hat{p}_{\phi 0} \hat{p}_{\phi 1} + 2 g^{\phi t} M \hat{E}_0 \hat{p}_{\phi 0} \nu^2 - g^{tt} \hat{E}_0^2 \nu^2 \\
& - g^{\phi\phi} M^2 \hat{p}_{\phi 0}^2 \nu^2.
\end{aligned}
\end{equation}
After each energy extraction, the remaining mass and angular momentum of the black hole are expressed as
\begin{equation}
M_n = M_{n-1} + \Delta M_{n-1} = M_{n-1} + \hat{E}_{1,n-1} \mu_{1,n-1}, \quad 
L_n = L_{n-1} + \hat{p}_{\phi 1} \mu_{1,n-1} M_{n-1},
\end{equation}
where 
\begin{equation}
L_0 = \hat{a}_0 M_0^2.
\end{equation}
This leads to corresponding changes as $\hat{a}=\frac{a}{M}$ and $\hat{N_s}$, namely
\begin{eqnarray}
\Delta \hat{a}_{n-1}=\frac{L_n}{M_n^{2}}-\frac{L_{n-1}}{M_{n-1}^2}, \quad \quad \Delta \hat{N_s}_{n-1}=\frac{N_s}{M_n^{2-\zeta}}-\frac{N_s}{M_{n-1}^{2-\zeta}}.
\end{eqnarray}
Crucially, the repetitive Penrose process is modelled under the assumption of a constant Rastall structure parameter $N_s$. This implies that the intrinsic deviation of the black hole from the Kerr metric remains invariant during energy extraction. Consequently, the dimensionless Rastall structure parameter $\hat{N_s}_{n}=\frac{N_s}{M_n^{2-\zeta}}$ was iteratively updated to reflect changes in mass. Similarly, both the radius of the horizon $r_+$ and the irreducible mass $M_{\text{irr}}$ evolve according to their respective governing equations. The resulting variation in the extractable energy is given by
\begin{equation}
\Delta E_{\mathrm{extractable},n-1} = \Delta M_{n-1} - \Delta M_{\mathrm{irr},n-1}.
\end{equation}
Consequently, the total extracted energy during the process is
\begin{equation}
E_{\mathrm{extracted},n} = M_0 - M_n.
\end{equation}
The energy return on investment $\xi$, which is characterized as the ratio of harvested energy to the total energy of all incident particles, which is defined as \cite{Lhu2026} 
\begin{equation}
\xi_n = \frac{E_{\mathrm{extracted},n}}{nE_0}.
\end{equation}
The energy utilization efficiency $\Xi_n$ is defined as the ratio of the harvested energy to the total reduction in the extractable energy of the black hole is expressed as \cite{Lhu2026} 
\begin{equation}
\Xi_n = \frac{E_{\mathrm{extracted},n}}{E_{\mathrm{extractable},0} - E_{\mathrm{extractable},n}}.
\end{equation}

\section{Iterative STOPPING Constraints}\label{sec3}

For the successful execution of the repetitive energy extraction process, it should satisfy certain conditions as described in \cite{Ruffini2025a, Lhu2026, Wang2025, Zeng2026}. For instance, initially, the mass deficit must obey the constraint as
\begin{equation}
1-\bar{\mu}_1-\bar{\mu}_2 > 0.
\end{equation}
Second, the energy of particle $1$ must remain strictly negative throughout the process, which requires $\hat{E}_1 < 0$. Third, the net extractable energy at the $n^{th}$ iteration step must be positive, such that $E_{\mathrm{extractable},n} > 0$. Fourth,  for each iteration the irreducible mass of the black hole must satisfy $\Delta M_{\mathrm{irr}} \ge 0$. Since $M_{\mathrm{irr}}$ remains invariant under reversible transformations and increases strictly during irreversible processes, a local decrease in the irreducible mass is physically forbidden, as it would constitute a direct violation of the generalized second law of thermodynamics via an entropy reduction. The final condition is that the radial turning points for particles $0$ and $2$ must be lie to the right of the peak value of effective potentials, whereas the turning point for particle $1$ must lie on the left side of the peak of its effective potential. The corresponding limiting case is when the classical turning point of each particle coincides precisely with the peak of its respective effective potential, such as
\begin{equation}
\hat{V}_i^+(\hat{r}_p) = \hat{E}_i, \quad \left. \frac{d\hat{V}_i^+}{d\hat{r}} \right|_{\hat{r}=\hat{r}_p} = 0,
\end{equation}
in which $\hat{r}_p = r_p/M$ denotes the dimensionless decay radius. In the specific case where $\hat{E}_0 = 1$, the minimum spin lower limit of the black hole required to terminate the iterative process at this point is controlled by particle $0$. This minimum spin lower limit corresponds precisely to the co-rotating marginally bound orbit of particle $0$. Furthermore, the angular velocity of a test particle restricted to a co-rotating Keplerian orbit within the background of a Rastall rotating black hole is given by
\begin{equation}
\Omega_K=\frac{aX+\sqrt{-2 ~r~X}}{2~r+a^2~X},
\end{equation}
where $X=\frac{-2 M}{r^2}+(\zeta-2)N_s r^{\zeta-3}$. Consequently, the specific energy of a particle in a co-rotating Keplerian orbit is given by 
\begin{equation}\label{g1}
\hat{\mathcal{E}} = -\frac{g_{tt} + g_{t\phi}\Omega_K}{\sqrt{-g_{tt} - 2g_{t\phi}\Omega_K-g_{\phi\phi}\Omega_K^2}}.
\end{equation}
For a marginally bound orbit where $\hat{\mathcal{E}}=1$, the minimum spin lower limit for particle $0$ is determined directly by solving Eq. (\ref{g1}). In contrast, when $\hat{E}_0 > 1$, this lower limit for halting the iteration is governed instead by particle $2$, with its corresponding minimum spin boundary situated precisely at the co-rotating photon sphere radius. The co-rotating photon sphere radius in the Rastall rotating black hole satisfies
\begin{equation}\label{eq:photon}
\left(2r-2M-\xi N_{s}r^{\xi-1}\right)^{2}-16r^{2}\left(r^{2}-2Mr+a^{2}-N_{s}r^{\xi}\right)=0.
\end{equation}
The lower boundary for the minimum spin of particle $2$ is extracted by solving Eq. (\ref{eq:photon}). To ensure the existence of a turning point for particle 1, the discriminant within the radical of Eq. (\ref{eq:muq1}) must remain positive. Therefore, to fulfill this requirement, the value of $\hat{r}_p$ should be greater than the radius of the event horizon of the black hole, while the critical case being $\hat{r}_p=\hat{r}_{+}$.

\begin{figure}[htbp]
\centering
\subfigure[\tiny][~Particle 0]{\label{2a}\includegraphics[width=5.3cm,height=5.1cm]{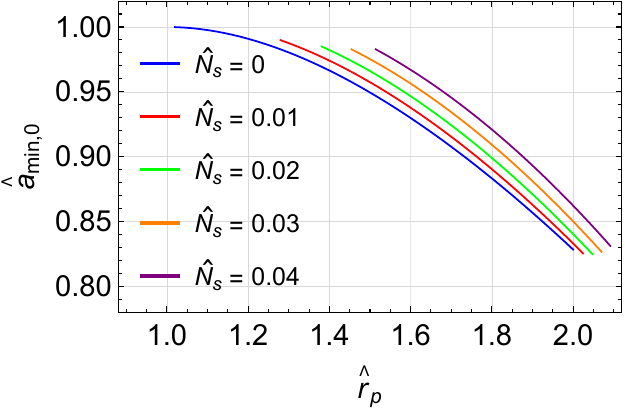}}
\subfigure[\tiny][~Particle 1]{\label{2b}\includegraphics[width=5.3cm,height=5.2cm]{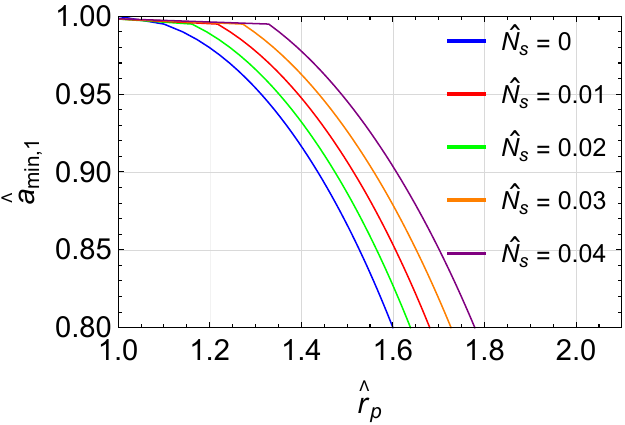}}
\subfigure[\tiny][~Particle 2]{\label{2c}\includegraphics[width=5.3cm,height=5cm]{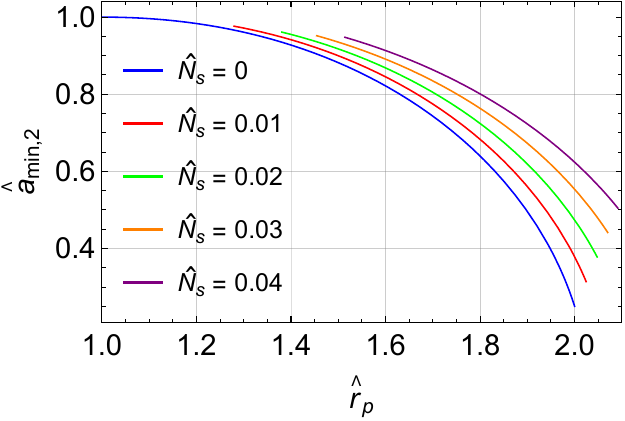}}
\subfigure[\tiny][~Particle 0]{\label{2d}\includegraphics[width=5.3cm,height=5cm]{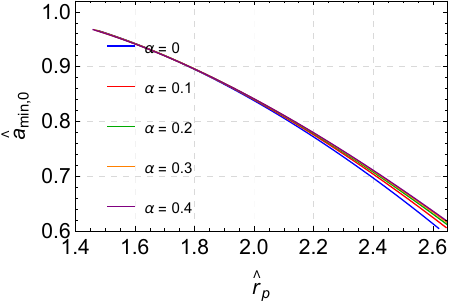}}
\subfigure[\tiny][~Particle 1]{\label{2e}\includegraphics[width=5.3cm,height=5cm]{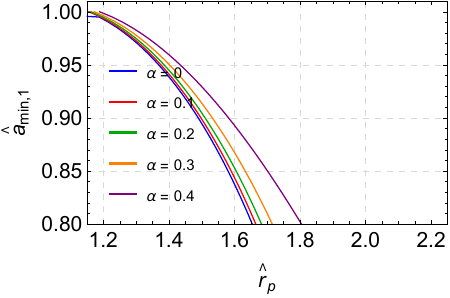}}
\subfigure[\tiny][~Particle 2]{\label{2f}\includegraphics[width=5.3cm,height=5cm]{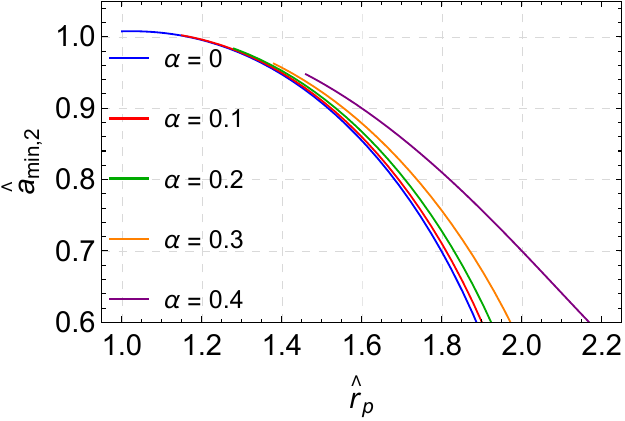}}
\caption{Variation of the minimum spin lower limits with the decay radius $\hat{r}_p$ for Particle $0$, Particle $1$ and Particle $2$ for different values of $\hat{N}_s$ with fixed $\alpha=0.05$ (top row), and for different values of $\alpha$ with fixed $\hat{N}_s=0.015$ (bottom row).}\label{fig2}
\end{figure}

In \textbf{FIG.}~\ref{fig2}, we analyze the dependency of the minimum spin lower limits for  Particle $0$, Particle $1$ and Particle $2$ with respect to $\hat{r}_p$ for different values of the parameters $\hat{N}_s$ and $\alpha$. It can be observe that from \textbf{FIG.}~\ref{fig2}, with the variations of both $\hat{N}_s$ and $\alpha$, the size of the ergosphere changes, so the range of the decay radius varies accordingly. For all three particles, the decay radius increase and the minimum spin lower limits decrease. Increasing the parameter values shifts the curves upward and outward. At a fixed decay radius, the lower spin limits of all three particles increase with increasing $\hat{N}_s$ and $\alpha$.

\begin{figure}[htbp]
\centering
\subfigure[\tiny][~$\hat{N}_s=0$,~$\alpha=0.05$]{\label{3a}\includegraphics[width=5.3cm,height=5.4cm]{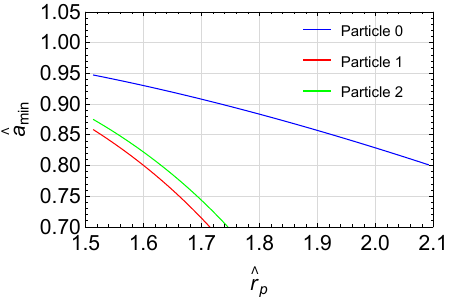}}
\subfigure[\tiny][~$\hat{N}_s=0.05$,~$\alpha=0.05$]{\label{3b}\includegraphics[width=5.3cm,height=5.4cm]{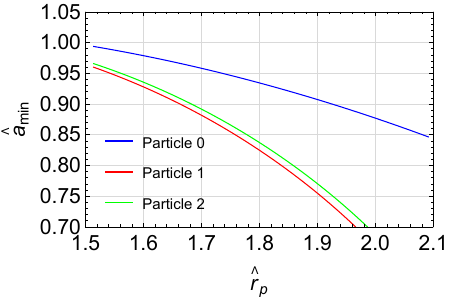}}
\subfigure[\tiny][~$\hat{N}_s=0.07$,~$\alpha=0.05$]{\label{3c}\includegraphics[width=5.3cm,height=5.4cm]{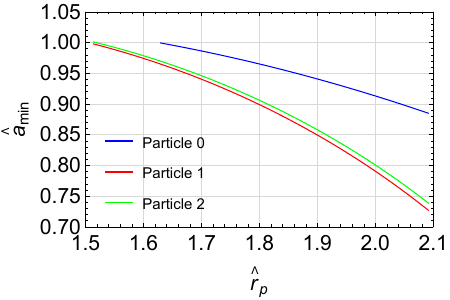}}
\centering
\subfigure[\tiny][~$\alpha=0$,~$\hat{N}_s=0.01$]{\label{3d}\includegraphics[width=5.3cm,height=5.4cm]{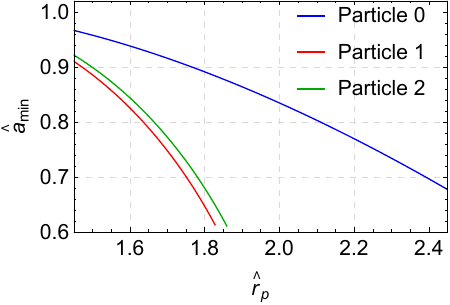}}
\subfigure[\tiny][~$\alpha=0.5$,~$\hat{N}_s=0.01$]{\label{3e}\includegraphics[width=5.3cm,height=5.4cm]{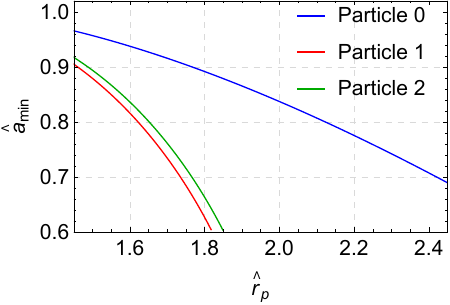}}
\subfigure[\tiny][~$\alpha=0.8$,~$\hat{N}_s=0.01$]{\label{3f}\includegraphics[width=5.3cm,height=5.4cm]{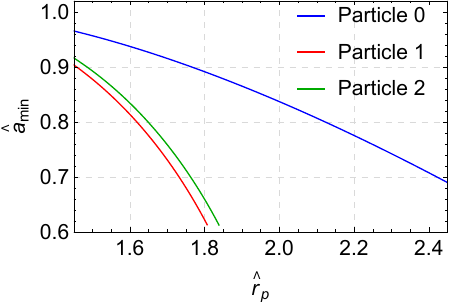}}
\caption{Comparison of the minimum spin lower limits for the three particles for different values of $\hat{N}_s$ and $\alpha$.}\label{fig3}
\end{figure}
In \textbf{FIG.}~\ref{fig3}, we illustrate the relative magnitudes of the minimum spin lower limits for the three particles for different values of both $\hat{N}_s$ and $\alpha$. It can be seen from the numerical results that a strict inequality holds in all configurations, namely $\hat{a}_{\text{min},1}<\hat{a}_{\text{min},2}<\hat{a}_{\text{min},0}$, and the largest of these values defines the spin lower limit, $a_{\min}=\max\left\{\hat{a}_{\min,0},\,\hat{a}_{\min,1},\,\hat{a}_{\min,2}\right\}$. Consequently, the spin lower limit required for stopping the iterative energy extraction process is uniquely controlled by Particle~0, which consistently demands the highest threshold. It should be emphasized that this spin lower limit for stopping the iteration is not fixed. In a dynamic extraction scenario, as energy is removed from the system, the black hole parameters shift, modifying the effective strength of the surrounding field parameter $\hat{N}_s$ and the Rastall coupling parameter $\alpha$.

From the first row of \textbf{FIG.}~\ref{fig3}, it is observed that as $\hat{N}_s$ increases from left to right, the corresponding curves gradually shift upward, resulting in a slight increase in the minimum spin required to stop the iterative process. In contrast, the second row of \textbf{FIG.}~\ref{fig3} illustrates the influence of the Rastall coupling parameter $\alpha$. As $\alpha$ increases, the curves remain nearly unchanged, indicating that variations in $\alpha$ have only a negligible effect on the stopping spin threshold. Although $\alpha$ modifies the horizon structure, the final minimum spin at which the iterative process stops remains predominantly sensitive to the Rastall structure parameter $\hat{N}_s$. This behavior is consistent with the corresponding effective potentials of these spacetimes, demonstrating that the final stage of the iterative energy extraction process is governed primarily by the physical constraints acting on Particle~$0$.

\section{The Repetitive Penrose Process In The Rastall Rotating Black Hole}\label{sec4}

Now, we are going to investigate the dynamics of the repetitive Penrose process. In this perspective, the authors in \cite{rufninew1} interpret that in the case of $\hat{E}_0>1$, the energy return on investment is lower than in the case of $\hat{E}_0=1$. Therefore, we choose $\hat{E}_0=1$ to maximize the energy return on investment. Closely followed by \cite{rufninew1}, we consider $\hat{p}_{\phi1}=-19.434$, $\nu=\frac{\mu_2}{\mu_1}=0.78345$ and we take $\mu_0=0.01M$ for conveniently.

\begin{table}[htbp]
\centering
\caption{The repetitive Penrose process for $\hat{N_s}=0.03$ , $\alpha=0.15$ and $\hat{r}_p=1.5$. The row highlighted in red represents the terminal iteration state of the repetitive Penrose process.}
\begin{tabular}{|p{0.3cm}| p{1.3cm}| p{1.3cm}| p{1.35cm}| p{1.65cm}| p{1.5cm}| p{1.5cm}| p{1.3cm}| p{1.5cm}| p{1.5cm}| p{1.75cm}| p{1.3cm}|}
  \hline
  \hline
~\textbf{n} &~~$\frac{M}{M_0}$ &~~~ $a$ &~~$\frac{M_{irr}}{M_0}$  & ~~$\frac{E_{extracted}}{M_0}$ &$\frac{E_{extractable}}{M_0}$&~~~~~$\xi$&~~~~$\Xi$&~~~$\frac{\mu_1}{\mu_0}$&~~~$\frac{\mu_2}{\mu_0}$&~~~$\frac{E_1}{\mu_0}$&$a_{critical}$\\
 
  \hline
  \hline

0 & 1.000000 & 0.990000 & 0.822557 & 0.000000 & 0.177443 & 0.000000 & 0.000000 & 0.000000 & 0.000000 & 0.000000 & 0.000000 \\
1 & 0.999112 & 0.987869 & 0.825055 & 0.00088844 & 0.174057 & 0.0888446 & 0.262361 & 0.0199946 & 0.0156647 & -0.0888446 & 0.976754 \\
2 & 0.998223 & 0.985753 & 0.827395 & 0.00177678 & 0.170828 & 0.0888391 & 0.268615 & 0.019885 & 0.0155789 & -0.0888336 & 0.97671 \\
3 & 0.997335 & 0.983653 & 0.829592 & 0.00266501 & 0.167743 & 0.0888336 & 0.274734 & 0.0197749 & 0.0154926 & -0.0888225 & 0.976666 \\
4 & 0.996447 & 0.981569 & 0.83166 & 0.00355312 & 0.164787 & 0.088828 & 0.280733 & 0.0196643 & 0.015406 & -0.0888115 & 0.976622 \\
5 & 0.995559 & 0.9795 & 0.83361 & 0.00444113 & 0.161949 & 0.0888225 & 0.286628 & 0.019553 & 0.0153188 & -0.0888004 & 0.976578 \\
6 & 0.994671 & 0.977448 & 0.835451 & 0.00532902 & 0.15922 & 0.088817 & 0.29243 & 0.0194411 & 0.0152311 & -0.0887894 & 0.976535 \\
\textcolor{red}{7} & \textcolor{red}{0.993783} & \textcolor{red}{0.975412} & \textcolor{red}{0.837191} & \textcolor{red}{0.0062168} & \textcolor{red}{0.156592} & \textcolor{red}{0.0888115} & \textcolor{red}{0.298151} & \textcolor{red}{0.0193285} & \textcolor{red}{0.0151429} & -\textcolor{red}{0.0887784} & \textcolor{red}{0.976491} \\
  \hline
  \hline
\end{tabular}\label{t1}
\end{table}

The results of this analysis are provided in \textbf{TABLE} \ref{t1}, where the Rastall structure parameter $\hat{N_s}=0.03$ and the Rastall parameter $\alpha=0.15$ are taken. In this case, the ergosphere is located at $(1.2806,2.0477)$ and we take the decay radius as $\hat{r}_p=1.5$. All data in \textbf{TABLE} \ref{t1} satisfied the iteration conditions. For instance, applying the mass deficit relation (25) yields $\tilde{\mu}_1 < 1/(1 + \mu_2/\mu_1) = 0.56$. Additionally, the conditions $\hat{E}_1 < 0$ and $E_{\mathrm{extractable},n} > 0$ are satisfied at each iteration, while the irreducible mass remains monotonically non-decreasing. The final column lists the minimum spin lower bound required to terminate the iterative process, demonstrating that this threshold marginally increases with successive iterations. At $n=7$ the iteration has stopped; if we forcibly further proceed to $8th$ iteration, we have $a_{8}<a_\text{critical,8}$ which no longer satisfy iteration conditions. As demonstrated in \textbf{TABLE}~\ref{t1}, a substantial fraction of the decrease in extractable energy is successfully converted into net extracted energy, while the remainder contributes to an increase in irreducible mass. Based on the calculated energy utilization efficiency $\Xi$, 29.81\% of the variation in extractable energy is recovered as extracted energy, while 70.19\% is converted into irreducible mass. These results indicate that reducing a black hole's spin cannot extract all its rotational energy, a limitation arising from nonlinear growth of its irreducible mass. Moreover, after the completion of the repetitive Penrose process, the remaining extractable energy settles at $0.1565926M$. This suggests that a significant amount of energy remains unharvested, which could be extracted by some other means. These results show resemblances with the case of Kerr black holes \cite{rufninew1}.

Next, we change the decay radius to $\hat{r}_p=1.7$, the Rastall structure parameter to $\hat{N}_s=0.07$ and the Rastall parameter to $\alpha=0.01$. The results are presented in \textbf{TABLE} \ref{t2} . From \textbf{TABLE} \ref{t2}, it can be seen that  iteration stops at $n=4$, causing  $\Xi$ to decrease at $24.47\%$ and the extractable energy reduces to $0.082911M$. Next, we change the decay radius to $\hat{r}_p=1.9$, keeping other parameters same as in the \textbf{TABLE} \ref{t2}. According to \textbf{TABLE} \ref{t3}, the iteration stops at $n=18th$ step. If we forcible proceed to $19th$ iteration, we would have $a_{19}<a_{\text{critical},19}$ which violates the iteration condition. From \textbf{TABLE} \ref{t3}, it can be seen that $\Xi$ decreases to $14.59\%$, $M_\text{irr}/M_0$ increase to $0.952648M$, leaving very small amount of extractable energy. The results shown in \textbf{TABLE} \ref{t3} are different from Kerr black holes , because after the termination of iteration there still exists a relative large amount of extractable energy, typically not less than $0.1M$ \cite{rufninew1}.

\begin{table}[htbp]
\centering
\caption{The repetitive Penrose process for $\hat{N_s}=0.07$ , $\alpha=0.01$ and $\hat{r}_p=1.7$}
\begin{tabular}{|p{0.3cm}| p{1.3cm}| p{1.3cm}| p{1.3cm}| p{1.8cm}| p{1.5cm}| p{1.5cm}| p{1.3cm}| p{1.5cm}| p{1.5cm}| p{1.7cm}| p{1.3cm}|}
  \hline
  \hline
\textbf{n} &$\frac{M}{M_0}$ & $a$ &$\frac{M_{irr}}{M_0}$  & $\frac{E_{extracted}}{M_0}$ &$\frac{E_{extractable}}{M_0}$&$\xi$& $\Xi$&$\frac{\mu_1}{\mu_0}$&$\frac{\mu_2}{\mu_0}$&$\frac{E_1}{\mu_0}$&$a_{critical}$\\
 
  \hline
  \hline

0 & 1.000000 & 0.990000 & 0.905800 & 0.0000000 & 0.0941996 & 0.000000 & 0.000000 & 0.000000 & 0.000000 & 0.000000 & 0.000000 \\
1 & 0.999309 & 0.987564 & 0.908073 & 0.000690918 & 0.0912358 & 0.0690918 & 0.233117 & 0.0195553 & 0.0153206 & -0.0690918 & 0.982287 \\
2 & 0.998618 & 0.985145 & 0.910248 & 0.0013817 & 0.0883703 & 0.0690849 & 0.237026 & 0.0194373 & 0.0152282 & -0.0690781 & 0.982218 \\
3 & 0.997928 & 0.982274 & 0.912331 & 0.00207234 & 0.0855971 & 0.0690781 & 0.240901 & 0.0193189 & 0.0151354 & -0.0690643 & 0.982149 \\
\textcolor{red}{4} & \textcolor{red}{0.997237} & \textcolor{red}{0.980362} & \textcolor{red}{0.914326} & \textcolor{red}{0.00276285} & \textcolor{red}{0.082911} & \textcolor{red}{0.0690712} & \textcolor{red}{0.244747} & \textcolor{red}{0.0191999} & \textcolor{red}{0.0150422} & -\textcolor{red}{0.0690505} & \textcolor{red}{0.98208} \\
  \hline
  \hline
\end{tabular}\label{t2}
\end{table}

\begin{table}[htbp]
\centering
\caption{The repetitive Penrose process for $\hat{N_s}=0.07$ , $\alpha=0.01$ and $\hat{r}_p=1.9$}
\begin{tabular}{|p{0.35cm}| p{1.3cm}| p{1.3cm}| p{1.3cm}| p{1.85cm}| p{1.5cm}| p{1.5cm}| p{1.3cm}| p{1.5cm}| p{1.5cm}| p{1.65cm}| p{1.3cm}|}
  \hline
  \hline
\textbf{n} &$\frac{M}{M_0}$ & $a$ &$\frac{M_{irr}}{M_0}$  & $\frac{E_{extracted}}{M_0}$ &$\frac{E_{extractable}}{M_0}$&$\xi$& $\Xi$&$\frac{\mu_1}{\mu_0}$&$\frac{\mu_2}{\mu_0}$&$\frac{E_1}{\mu_0}$&$a_{critical}$\\
 
  \hline
  \hline

0 & 1 & 0.99 & 0.9058 & 0 & 0.0941996 & 0 & 0 & 0 & 0 & 0 & 0 \\
1 & 0.999554 & 0.98677 & 0.909318 & 0.000445533 & 0.0902368 & 0.0445533 & 0.112431 & 0.021143 & 0.0165645 & -0.0445533 & 0.935858 \\
2 & 0.999109 & 0.983558 & 0.912684 & 0.000890977 & 0.0864255 & 0.0445488 & 0.114608 & 0.0210265 & 0.0164732 & -0.0445443 & 0.935805 \\
3 & 0.998664 & 0.980365 & 0.915910 & 0.00133633 & 0.0827532 & 0.0445443 & 0.116747 & 0.0209102 & 0.0163821 & -0.0445353 & 0.935752 \\
4 & 0.998218 & 0.97719 & 0.919009 & 0.00178159 & 0.0792094 & 0.0445398 & 0.118851 & 0.020794 & 0.0162911 & -0.0445262 & 0.935699 \\
5 & 0.997773 & 0.974032 & 0.921988 & 0.00222676 & 0.0757848 & 0.0445353 & 0.120923 & 0.020678 & 0.0162002 & -0.0445171 & 0.935646 \\
6 & 0.997328 & 0.970893 & 0.924857 & 0.00267184 & 0.0724713 & 0.0445307 & 0.122966 & 0.0205622 & 0.0161095 & -0.044508 & 0.935594 \\
7 & 0.996883 & 0.967773 & 0.927622 & 0.00311683 & 0.0692616 & 0.0445262 & 0.124983 & 0.0204466 & 0.0160189 & -0.0444988 & 0.935541 \\
8 & 0.996438 & 0.96467 & 0.930289 & 0.00356173 & 0.0661493 & 0.0445216 & 0.126977 & 0.020331 & 0.0159283 & -0.0444897 & 0.935488 \\
9 & 0.995993 & 0.961586 & 0.932865 & 0.00400653 & 0.0631286 & 0.044517 & 0.128948 & 0.0202156 & 0.0158379 & -0.0444804 & 0.935435 \\
10 & 0.995549 & 0.958519 & 0.935355 & 0.00445124 & 0.0601942 & 0.0445124 & 0.130898 & 0.0201004 & 0.0157476 & -0.0444712 & 0.935383 \\
11 & 0.995104 & 0.955471 & 0.937763 & 0.00489586 & 0.0573416 & 0.0445078 & 0.132831 & 0.0199852 & 0.0156574 & -0.0444619 & 0.93533 \\
12 & 0.99466 & 0.952441 & 0.940009 & 0.00534039 & 0.0545665 & 0.0445032 & 0.134746 & 0.0198702 & 0.0155673 & -0.0444526 & 0.935277 \\
13 & 0.994215 & 0.94943 & 0.94235 & 0.00578482 & 0.0518649 & 0.0444986 & 0.136645 & 0.0197553 & 0.0154773 & -0.0444432 & 0.935225 \\
14 & 0.993771 & 0.946437 & 0.944538 & 0.00622916 & 0.0492333 & 0.044494 & 0.13853 & 0.0196404 & 0.0153873 & -0.0444338 & 0.935172 \\
15 & 0.993327 & 0.943461 & 0.946658 & 0.0066734 & 0.0466686 & 0.0444894 & 0.140401 & 0.0195256 & 0.0152974 & -0.0444244 & 0.93512 \\
16 & 0.992882 & 0.940505 & 0.948715 & 0.00711755 & 0.0441676 & 0.0444847 & 0.14226 & 0.019411 & 0.0152075 & -0.044415 & 0.935067 \\
17 & 0.992438 & 0.937566 & 0.950711 & 0.00756161 & 0.0417276 & 0.04448 & 0.144108 & 0.0192963 & 0.0151177 & -0.0444055 & 0.935015 \\
\textcolor{red}{18} & \textcolor{red}{0.991994} & \textcolor{red}{0.934646} & \textcolor{red}{0.952648} & \textcolor{red}{0.00800557} & \textcolor{red}{0.0393461} & \textcolor{red}{0.0444754} & \textcolor{red}{0.145945} & \textcolor{red}{0.0191817} & \textcolor{red}{0.0150279} & -\textcolor{red}{0.044396} & \textcolor{red}{0.934962} \\
  \hline
  \hline
\end{tabular}\label{t3}
\end{table}

Finally, we take small decay radius $\hat{r}_p=1.3$, $\hat{N}_s=0.01$ and $\alpha=0.53$. The results are summarized in \textbf{TABLE} \ref{tp4}. The process terminates at $n=1st$ iteration which gives the optimal amount of the energy utilization energy $\Xi=43.68$ and the remaining amount of extractable energy is $0.20286M$. This is significantly larger than \cite{Ruffini2025a,Wang2025,Zeng2026,Zeng2,guass}. Further if the decay radius is small like $\hat{r}_p=1.29$ the iteration even stops at $n=0$, because then $a=0.99$ and $a_\text{critical}=0.991161$ which violates the iteration condition.

\begin{table}[htbp]
\centering
\caption{The repetitive Penrose process for $\hat{N}_s=0.01$ , $\alpha=0.52$ and $\hat{r}_p=1.3$}
\begin{tabular}{|p{0.3cm}| p{1.3cm}| p{1.3cm}| p{1.3cm}| p{1.7cm}| p{1.5cm}| p{1.5cm}| p{1.3cm}| p{1.3cm}| p{1.5cm}| p{1.5cm}| p{1.3cm}|}\hline\hline
\textbf{n} &~~~$\frac{M}{M_0}$ &~~~~ $a$ &~~$\frac{M_{irr}}{M_0}$  &~ $\frac{E_{extracted}}{M_0}$ &$\frac{E_{extractable}}{M_0}$&~~~~~$\xi$& ~~~~$\Xi$&~~~$\frac{\mu_1}{\mu_0}$&~~~~$\frac{\mu_2}{\mu_0}$&~~~$\frac{E_1}{\mu_0}$&~$a_{critical}$\\\hline\hline
0 & 1.000000 & 0.990000 & 0.794260 & 0.000000 & 0.205740 & 0.000000 & 0.000000 & 0.000000 & 0.000000 & 0.000000 & 0.000000 \\
\textcolor{red}{1} & \textcolor{red}{0.998742} & \textcolor{red}{0.988748} & \textcolor{red}{0.795882} & \textcolor{red}{0.00125845} & \textcolor{red}{0.20286} & \textcolor{red}{0.125845} & \textcolor{red}{0.436857} & \textcolor{red}{0.019239} & \textcolor{red}{0.0150728} & -\textcolor{red}{0.125845} & \textcolor{red}{0.989923} \\ \hline\hline
\end{tabular}\label{tp4}
\end{table}

\begin{figure}[htbp]
\centering
\subfigure[\tiny][~]{\label{4a}\includegraphics[width=7.2cm,height=6.6cm]{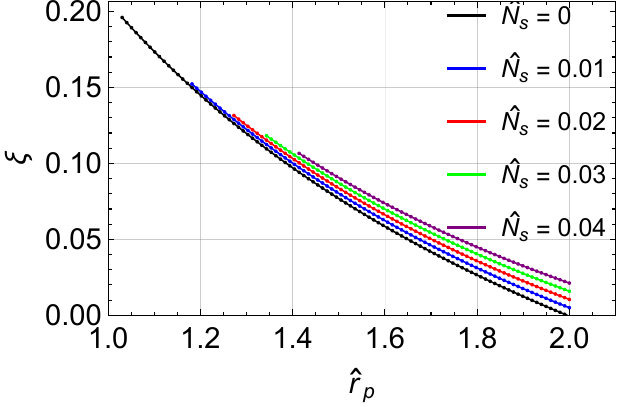}}\quad\quad
\subfigure[\tiny][~]{\label{4b}\includegraphics[width=7.2cm,height=6.6cm]{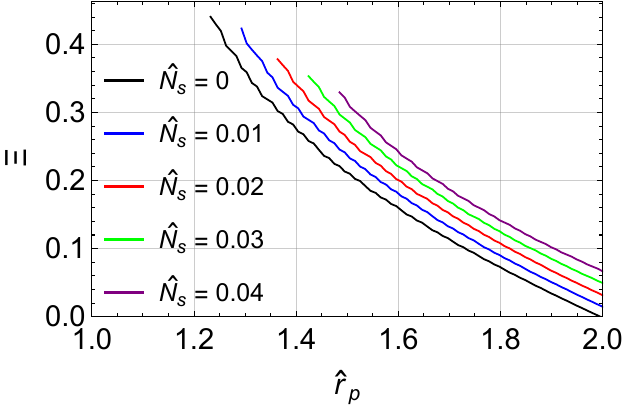}}

\subfigure[\tiny][~]{\label{4c}\includegraphics[width=7.2cm,height=6.6cm]{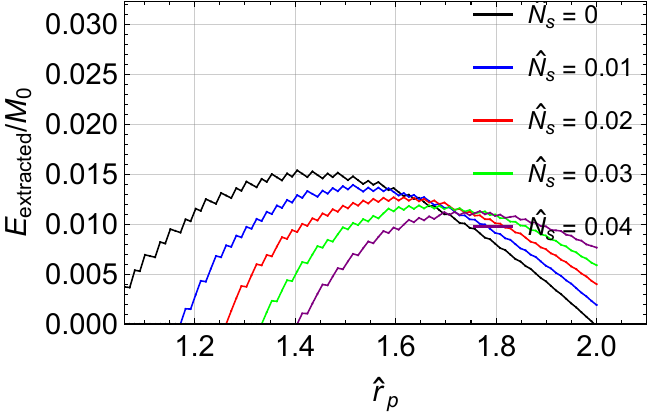}}\quad\quad
\subfigure[\tiny][~]{\label{4d}\includegraphics[width=7.2cm,height=6.6cm]{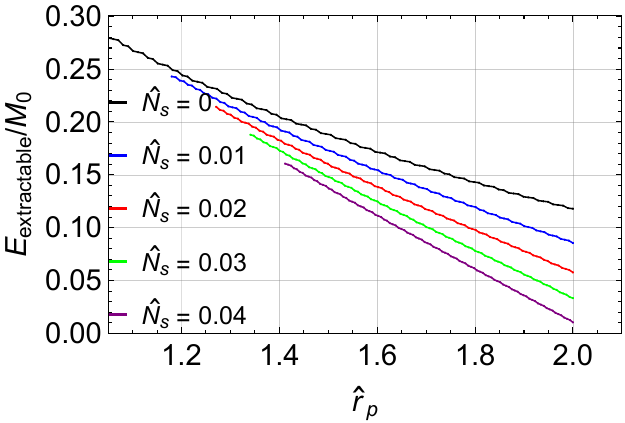}}
\caption{Panels (a)-(d) illustrate the variation of key energetic parameters with the decay radius $\hat{r}_p$ following the termination of the repetitive Penrose process, for different initial values of $\hat{N}_s$ with fixed $\alpha=0.05$. Specifically, panel (a) shows the energy return on investment, $\xi$; panel (b) presents the energy utilization efficiency, $\Xi$; panel (c) depicts the extracted energy, $E_{\mathrm{extracted}}/M_0$; and panel (d) illustrates the extractable energy, $E_{\mathrm{extractable}}/M_0$. The oscillatory behavior exhibited by these curves arises from changes in the number of iterations, which are governed by the iteration conditions, thereby reflecting the discrete nature of the repetitive Penrose process.}\label{fig4}
\end{figure}
\textbf{FIG.} \ref{fig4} presents the variation of the energy return on investment $\xi$, the energy utilization efficiency $\Xi$, the extracted energy $E_{\mathrm{extracted}}/M_0$ and the extractable energy $E_{\mathrm{extractable}}/M_0$ as a function of the decay radii $\hat{r}_p$ after the completion of the repetitive Penrose process for different values of Rastall structure parameter $\hat{N}_s$. The result shows that the surrounding field distribution has significant effects on the energy extraction process throughout the iterative decay process.

Particularly, panels (a) and (b) of \textbf{FIG.}~\ref{fig4} illustrates that under the same decay radius, the energy return is greater for each increasing value of the structure parameter $\hat{N}_s$. In addition, $\xi$ decreases monotonically with increasing decay radius $\hat{r}_p$, indicating that the contribution from successive particle returns gradually weakens as the decay radius increases. This behavior implies that the efficiency of the repetitive Penrose process progressively declines at larger decay radii, leading to diminishing energy gains from subsequent iterations. Further, greater values of $\hat{N}_s$ produce lower maximum values of energy return suggesting that a stronger surrounding field reduces the maximum amount of energy recovered during the initial stages of the repetitive Penrose process. The behavior of the energy utilization efficiency $\Xi$, reflects that at the same decay radii it increases gradually with increasing $\hat{N}_s$. The energy utilization efficiency decreases continuously with the aid of $\hat{r}_p$ and rises rapidly for lower values of radius $\hat{r}_p$. This trend shows that, the remaining energy of the black hole is converted into usefull extracted energy more effectively at the initial stages of repetitive Penrose process. In addition, increasing $\hat{N}_s$ shifts the curves upward indicating that the surrounding Rastall field boosts the energy conversion efficiency.

The bottom row of \textbf{FIG.}~\ref{fig4} (panels c and d) displays the effects of $\hat{N}_s$ on the extracted energy $E_{\mathrm{extracted}}/M_0$ and the extractable energy $E_{\mathrm{extractable}}/M_0$, respectively. From Fig.~ \ref{fig4} (c), it can be seen that the extracted energy exhibits the parabola, for larger values of $\hat{N}_s$, the peak values of extracted energy are smaller and the curves shift to higher decay radius $\hat{r}_p$. The parabolic curves reflect that the extracted energy first increases, reaches a distinct maximum and then gradually decreases. This is due to the balance between the increasing efficiency of individual iterations and continuous reduction in the available energy reservoir. As $\hat{N}_s$ increases, the location of the maximum extracted energy shifts toward higher decay radii, indicating that a stronger surrounding field delays the radius at which the maximum energy extraction is achieved. \textbf{FIG.} \ref{fig4} (c) shows that a larger initial values of $\hat{N}_s$ results in a smaller extractable energy. For all values of $\hat{N}_s$, the extractable energy decreases linearly with increasing decay radius $\hat{r}_p$. Its reason is the exhaustion of the black hole's rotational energy during the energy extraction process. Furthermore, greater values of $\hat{N}_s$ results in steeper decline in extractable energy implying that the available energy reservoir is consumed more rapidly in the presence of a stronger surrounding field. The small oscillations in curves shows the discrete nature of iterations. Since the energy extraction proceeds through iterations, rather than in continuous evolution therefore each completed iteration introduces a small change in energetic quantities. Also, the decay radii increase the number of iterations changes leading to oscillations in the corresponding curves.

\begin{figure}[htbp]
\centering
\subfigure[\tiny][~]{\label{5a}\includegraphics[width=7.2cm,height=6.6cm]{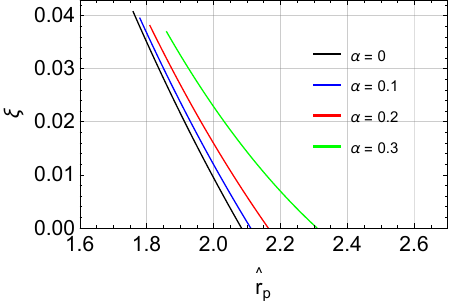}}\quad\quad
\subfigure[\tiny][~]{\label{5b}\includegraphics[width=7.2cm,height=6.6cm]{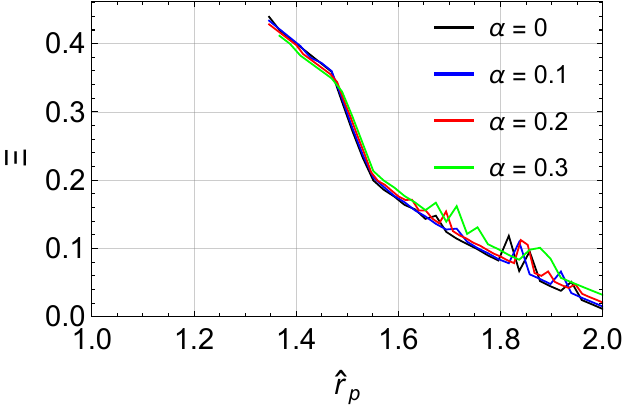}}
\subfigure[\tiny][~]{\label{5c}\includegraphics[width=7.2cm,height=6.6cm]{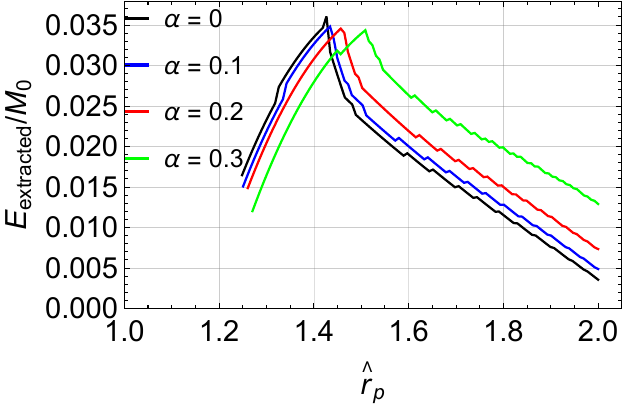}}\quad\quad
\subfigure[\tiny][~]{\label{5d}\includegraphics[width=7.2cm,height=6.6cm]{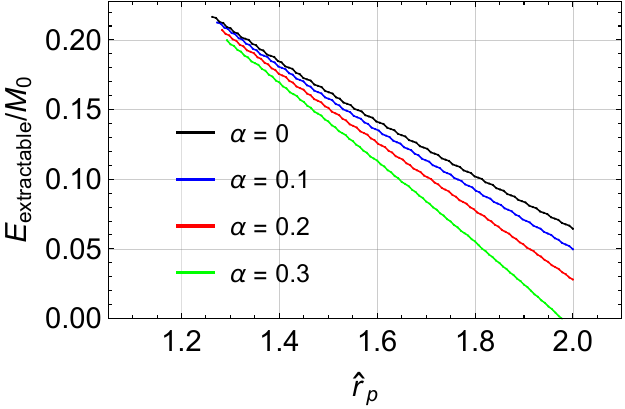}}
\caption{Variation of $\xi$, $\Xi$, $E_{\mathrm{extracted}}/M_0$ and  $E_{\mathrm{extractable}}/M_0$ for different initial values of $\alpha$ with fixed $\hat{N}_s=0.02$.}\label{fig5}
\end{figure}

\textbf{FIG.}~\ref{fig5} depicts the dependence of different values of the Rastall parameter on the energy return on investment $\xi$, the energy utilization efficiency $\Xi$, the extracted energy $E_{\mathrm{extracted}}/M_0$ and the extractable energy $E_{\mathrm{extractable}}/M_0$. The results demonstrate that varying $\alpha$ changes the efficiency of energy extraction and the evolution of the available energy. \textbf{FIG.}~\ref{fig5}(a) reflects that the energy return on investment decreases with increasing decay radius, and larger values of $\alpha$ result in a decreasing maximum value of $\xi$. It can be seen from \textbf{FIG.}~\ref{fig5}(b), that the energy utilization efficiency $\Xi$ exhibits an overall monotonic decrease with respect to the increasing decay radius $\hat{r}_p$. Within the small-radius regime, $\Xi$ converges asymptotically to virtually identical values for all considered choices of $\alpha$, suggesting that the non-minimal gravitational coupling effects are subdominant near the horizon interface. In contrast, for larger values of the decay radius, a clear stratification emerges: the energy utilization efficiency increases monotonically with the Rastall parameter $\alpha$. The extracted energy $E_{\mathrm{extracted}}/M_0$, shown in \textbf{FIG.}~\ref{fig5} (c), initially increases sharply as compared to that indicated in \textbf{FIG.}~\ref{fig4}, reaches a maximum value and then decreases gradually. The peak values of extracted energy decrease with increasing $\alpha$, shifts towards larger values of $\hat{r}_p$. This demonstrates that the $\alpha$ modifies the location and the evolution of the maximum energy extraction. The behavior of the extractable energy displayed in \textbf{FIG.}~\ref{fig5} (d) shows that the remaining extractable energy decreases with increasing decay radius. Larger values of $\alpha$ shift the maximum extractable energy values downward, implying a depletion of the available rotational energy reserve.

\begin{figure}[htbp]
\centering
\subfigure[\tiny][~$\hat{r}_p=1.3$]{\label{6a}\includegraphics[width=7.2cm,height=6.6cm]{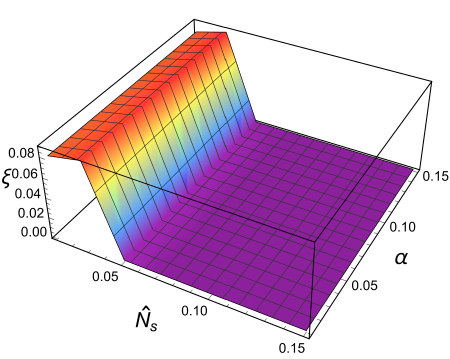}}\quad\quad
\subfigure[\tiny][~$\hat{r}_p=1.5$]{\label{6b}\includegraphics[width=7.2cm,height=6.6cm]{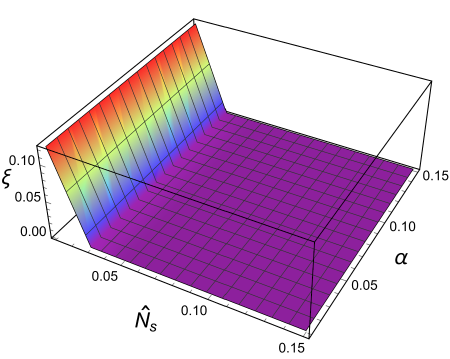}}

\subfigure[\tiny][~$\hat{r}_p=1.7$]{\label{6c}\includegraphics[width=7.2cm,height=6.6cm]{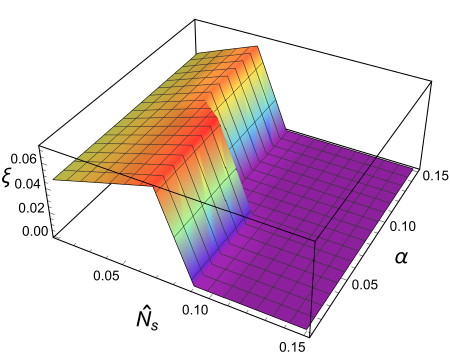}}\quad\quad
\subfigure[\tiny][~$\hat{r}_p=1.9$]{\label{6d}\includegraphics[width=7.2cm,height=6.6cm]{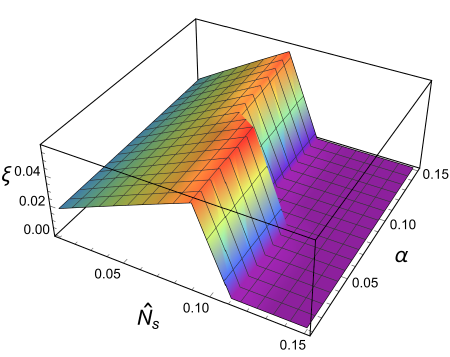}}
\caption{The maximum energy return on investment $\xi$ as a function of $\hat{N}_s$ and $\alpha$ for different values of decay radius $\hat{r}_p$.}\label{fig6}
\end{figure}

\textbf{FIG.}~\ref{fig6} presents three-dimensional display, depicting the behavior of the maximum energy return on investment $\xi$ over a parameter space defined by $\hat{N}_s$ and $\alpha$ for four distinct values of the decay radius $\hat{r}_p$. Across all panels, $\xi$ attains its highest values in the low $\hat{N}_s$ regime (approximately $\hat{N}_s \lesssim 0.05$), beyond which it decreases monotonically and asymptotically approaches negligible magnitudes. The sensitivity of $\xi$ with respect to $\alpha$ is strongly governed by the choice of $\hat{r}_p$. For $\hat{r}_p = 1.3$, $1.5$, and $1.7$,  panels~(a)-(c) reveal that the surface profiles remain nearly flat along $\alpha$ in the high $\xi$ region, reflecting minimal dependence of the maximum energy return on investment on $\alpha$. In contrast, panel (d) for $\hat{r}_p = 1.9$ exhibits a distinct behavior, where a well-defined ridge develops whose amplitude varies noticeably with $\alpha$, accompanied by a well-defined suppression region at intermediate  values of $\hat{N}_s$. Furthermore, a comparative inspection of the vertical axes across all panels reveals that the peak magnitude of $\xi$ progressively diminishes as $\hat{r}_p$ increases from $1.3$ to $1.9$, indicating that larger decay radii reduce the overall maximum energy return on investment while simultaneously enhancing the system's sensitivity to $\alpha$.

\begin{figure}[htbp]
\centering
\subfigure[\tiny][~$\hat{r}_p=1.3$]{\label{6a}\includegraphics[width=7.2cm,height=6.6cm]{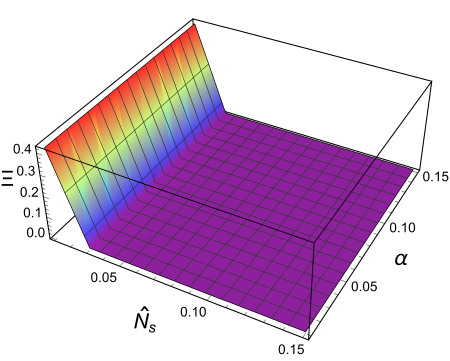}}\quad\quad
\subfigure[\tiny][~$\hat{r}_p=1.5$]{\label{6b}\includegraphics[width=7.2cm,height=6.6cm]{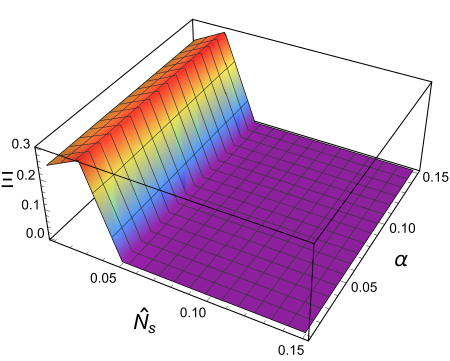}}
\subfigure[\tiny][~$\hat{r}_p=1.7$]{\label{6c}\includegraphics[width=7.2cm,height=6.6cm]{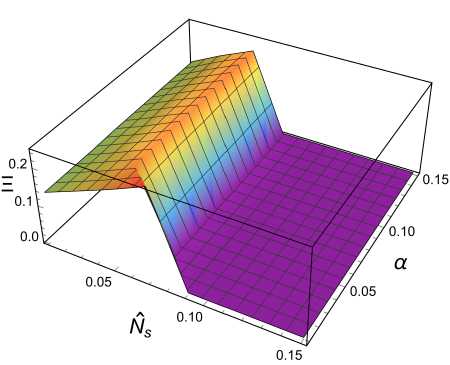}}\quad\quad
\subfigure[\tiny][~$\hat{r}_p=1.9$]{\label{6d}\includegraphics[width=7.2cm,height=6.6cm]{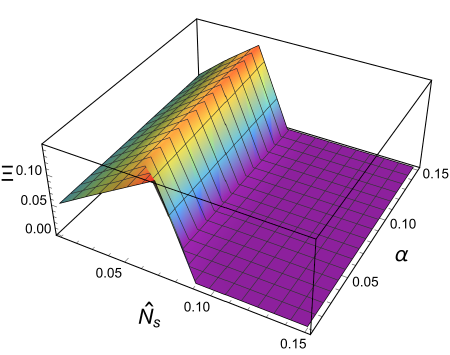}}
\caption{The maximum energy utilization efficiency $\Xi$ as a function of $\hat{N}_s$ and $\alpha$ for different values of decay radius $\hat{r}_p$.}\label{fig7}
\end{figure}
\textbf{FIG.}~\ref{fig7} illustrates the three-dimensional variation of the maximum energy utilization efficiency $\Xi$ as a function of $\hat{N}_s$ and $\alpha$ for four representative values of the decay radius $\hat{r}_p$. A globally consistent trend across all panels is the systematic suppression of peak $\Xi$ with increasing $\hat{r}_p$; the maximum value diminishes from approximately $0.4$ at $\hat{r}_p = 1.3$ to nearly $0.10$ at $\hat{r}_p = 1.9$, demonstrating that the energy extraction efficiency is a decreasing function of the decay radius. At smaller decay radii, $\hat{r}_p = 1.3$ and $1.5$ in panels (a) and (b), the maximum $\Xi$ is sharply concentrated within $\hat{N}_s \lesssim 0.05$ and exhibits negligible variation with respect to $\alpha$, indicating that the efficiency is predominantly governed by $\hat{N}_s$ alone, with the 
Rastall parameter $\alpha$ playing a subdominant role. In contrast, for $\hat{r}_p = 1.7$ and $1.9$ in panels~(c) and~(d), the surface transitions from a monotonically declining profile into one featuring a localized prominent ridge that shifts toward higher $\hat{N}_s$ values while acquiring a measurable dependence on $\alpha$. This structural evolution of the efficiency landscape demonstrates that maximizing $\Xi$ in the repeated Penrose process demands precise and simultaneous fine-tuning of both the black hole spacetime geometry and the energy extraction parameters, reflecting the inherently non-trivial optimization in modified gravity frameworks.

\section{Conclusion}\label{sec5}
In this work, we have investigated the repetitive Penrose process for neutral particles in a Rastall rotating black hole surrounded by a quintessence field, treating every decay self-consistently through the triple turning-point condition and updating the mass, angular momentum, and irreducible mass at each step. To verify the numerical consistency of our analysis, we consider the limiting case $\hat{N}_s=0$ and compare our results with those reported in \cite{rufninew1}. Specifically, we examined how the Rastall structure parameter and the Rastall coupling influence the extraction of rotational energy. First, we provided the theoretical framework for the repetitive Penrose process in the Rastall rotating black hole by deriving the conservation equations together with the corresponding iterative evolution equations for the event horizon, ergosphere, mass of the black hole, spin, irreducible mass, and extractable energy. Then, a complete set of physical termination conditions was formulated to determine the endpoint of the repetitive extraction process. Through an analysis of the minimum spin lower limits for the three decay particles, we showed that the stopping condition is always governed by Particle~$0$, whose critical spin threshold remains higher than those of the other particles throughout the parameter space considered. This analysis shows that the repetitive Penrose process gradually extracts rotational energy, due to which the irreducible mass of the black hole increases step by step. As a result the spin of the black hole decreases to a critical value, which makes a fraction of energy inaccessible through this mechanism. This confirms that the growth of the irreducible mass provides a fundamental limit to the efficiency of repetitive energy extraction, consistent with the generalized laws of black hole thermodynamics. The surrounding Rastall field parameter, $\hat{N}_s$, and the Rastall coupling parameter, $\alpha$, both significantly modify the energy extraction process. Increasing $\hat{N}_s$ elevates the energy utilization efficiency, shifts the maximum extracted energy toward larger decay radii, and cause the reduction in extractable energy, although the energy return on investment is marginally reduced. Similarly, larger values of $\alpha$ increase the utilization efficiency at higher decay radii shifts the maximum value of the extracted energy to higher decay radii and result in smaller maximum value of extracted energy. These results demonstrate that both the $\hat{N}_s$ and $\alpha$ significantly influence the dynamics and efficiency of the repetitive Penrose process. Our results indicate that modified gravity effects encoded in the Rastall rotating black hole leave noticeable impacts on the repetitive Penrose process. For better comparison with the existing literature, \textbf{TABLE}~\ref{tab2} summarizes the key differences between the repetitive Penrose process in the Rastall rotating black hole and previous studies conducted within the framework of other black hole models.

\begin{table}[htbp]
\centering
\caption{Comparison of the repetitive Penrose process in different black hole spacetimes. $M_{\text{irr}}$ is the irreducible mass, EUE is the energy utilization efficiency and EROI is the energy return on investment.}
\label{tab2}
\begin{tabular}{|p{2cm}| p{2.5cm}| p{2.1cm}| p{1.5cm}| p{6.6cm}| p{1.0cm}|}
  \hline
  \hline
\textbf{Black Holes} & \textbf{~~~Modifiers} & \textbf{Governing Parameters} & \textbf{~EUE $ \%$} & \textbf{~~~~~~~~~Physical Interpretation}&\textbf{~Ref.}\\
  \hline
  \hline
  
Kerr &Spin &decay radius $\hat{r}$&~~$25.32\%$&The majority of rotational energy is transformed into $M_{\text{irr}}$; meaning spin cannot be fully depleted.&~~\cite{rufninew1}\\
\hline

Reissner-Nordstr\"om&Charge &$\hat{q}_1$, $\hat{r}$&~~$21.8\%$&Charge cannot be entirely neutralize and EUE is bounded below $50\%$.&~~\cite{Lhu2026}\\
\hline

Kerr-de ~~~Sitter &Spin, $\Lambda$ &$\Lambda M^2$&~~$24.05\%$&Positive cosmological constant enhances the EROI but EUE $<50\%$.&~~\cite{Wang2025}\\
\hline

Konoplya-Zhidenko & Spin, deformation parameter $\eta$ &~~~$\hat{\eta}$, $\hat{r}$&~~$23.93\%$& At same decay radius deformation parameter $\hat{\eta}$ increases both EROI and EUE but EUE$<50\%$.& ~~\cite{Zeng2}\\
\hline

Rastall Rotating & Spin, structure field parameter $N_s$, and coupling parameter $\alpha$ & $\hat{N}_s$, $\alpha$ and $\hat{r}$ &~~$43.69\%$& Rastall field and Rastall gravity parameters boosts EROI at same decay radius and maximum EROI at smaller $\hat{r}$. The larger values of both these parameters enhances EUE rapidly at higher decay radius resulting EUE$>50\%$.& This work\\
\hline
\hline
\end{tabular}
\end{table}

\end{document}